# Time-Indexed Parallel Spatial Modulation for Large-scale MIMO Systems with Antenna Grouping

Taissir Y. Elganimi* and Khaled M. Rabie^Δ
*Department of Electrical and Electronic Engineering, University of Tripoli, Libya
^ΔDepartment of Engineering, Manchester Metropolitan University, United Kingdom
Emails: t.elganimi@uot.edu.ly and k.rabie@mmu.ac.uk

*Abstract*—A novel space-time parallel index modulation scheme is proposed in this paper for conveying extra digital information with the aid of space and time indices. In this proposed scheme, time-indexing is cleverly performed in parallel spatial modulation (PSM) schemes using transmit antenna grouping, and referred to as time-indexed parallel spatial modulation (TI-PSM). In this scheme, transmit antennas are divided into groups to adopt TI-SM scheme in large-scale multiple input multiple output (MIMO) systems with inter-channel interference (ICI) avoidance. This scheme is attractive due to both the high data rate and good performance improvement that can be achieved. The bit error rate (BER) performance of the TI-PSM scheme is evaluated and compared to that of the conventional schemes. Simulation results demonstrated that a significant improvement is achieved by the TI-PSM scheme as compared to the TI-SM and PSM schemes for the same achieved rate. It is also shown that TI-SM and TI-PSM schemes provide higher performance drop with channel estimation errors (CEEs) than the conventional SM and PSM systems. Therefore, due to the high performance improvements achieved in TI-PSM scheme, it can be effectively regarded as a promising solution for various 5G and beyond wireless networks.

## I. INTRODUCTION

Conventional spatial modulation (SM) scheme is regarded as the most popular index modulation (IM) transmission technique in the literature because of its high data rate, low cost and power consumption, and low receiver complexity. In SM, information bits are conveyed by performing symbol transmission in spatial domain besides the information bits that are conveyed by conventional complex modulation alphabets such as $M$-ary quadrature amplitude modulation (QAM) and phase shift keying (PSK) [1, 2]. However, multidimensional IM schemes have recently caught the attention of many researchers as innovative techniques in multiple input multiple output (MIMO) communications where new dimensions are provided to convey additional information bits by utilizing the indices of multiple transmission resources in domains other than the spatial domain. For example, in [3], the indices of time slots in the time domain are exploited to convey extra information bits, in [4], the indices of the subcarriers in the frequency domain in multicarrier systems are utilized and termed as subcarrier index modulation (SIM), while the indices of the radio frequency (RF) mirrors are used in [5, 6]. Furthermore, several IM schemes have been considered as alternative solutions for the fifth generation (5G) and beyond wireless systems because of their significant performance improvements [7, and references therein].

Space-time IM introduced in [3] is a novel IM scheme in which slot indexing is performed efficiently in time and spatial domains. This scheme is also termed in [12] as time-indexed spatial modulation (TI-SM), where transmission has been carried out in frames, each consists of a certain number of several time slots. During transmission, it is not necessary for all time slots in a data frame to be activated. In other words, no transmission occurs in a certain time slots since a fraction of the time slots are used for transmission, and the corresponding indices are utilized to carry information. Moreover, on the used time slots, transmission of modulation symbols from a chosen transmit antenna occurs based on the antenna index bits [3]. Therefore, information bits in TI-SM scheme are conveyed through indexing in time and spatial domains besides the indices of the modulation symbols (i.e., $\log_2 M$ bits), where $M$ is the constellation size.

Recently, several research studies have focused on transmit antenna grouping in SM schemes in order to increase the number of bits conveyed over the spatial domain. In [8] and [9], the available transmit antennas of SM and generalized spatial modulation (GSM) schemes, respectively, are equally divided into groups. Each group, however, independently performs the conventional antenna indexing scheme with the same signal symbol. Later, the scheme in [9] is extended in another research work to massive MIMO systems [10]. More recently, parallel quadrature spatial modulation (PQSM) is proposed in [11] by adopting quadrature spatial modulation (QSM) scheme in massive MIMO systems and dividing the transmit antennas into groups. All these schemes use only one RF chain which is a key advantage of SM schemes.

In this paper, a new IM scheme is proposed, and referred to as *time-indexed parallel spatial modulation (TI-PSM)*. It can be also termed as *space-time parallel index modulation (STPIM)*. This scheme implies dividing the transmit antennas into groups, each of which includes an even number of antennas that has to be a power of two. From each group, only one transmit antenna is activated at any particular time instant to emit the same modulated symbol from all groups in parallel. In TI-PSM scheme, conveying extra information bits through time-slot indexing is based on the choice of the used and unused time slots in a data frame, where the choice of the used slots conveys slot index bits. Furthermore, in parallel spatial modulation (PSM) scheme that considered in this paper using transmit antenna grouping, it is not necessary that the number of groups and the total number of transmit antennas to be powers of two, while the number of transmit antennas in each group has to be a power of two in order to perform the conventional SM scheme in each group. This paper has also studied the effect of channel estimation errors (CEEs) on the error performance of all schemes.

The proposed TI-PSM scheme presents attractive and considerable features including, *i)* TI-PSM achieves higher spectral efficiency than TI-SM scheme if the same configuration and modulation order are used; *ii)* TI-PSM achieves the best tradeoff between the orders of the spatial modulation and the signal constellation, thus enhancing the bit error rate (BER) performance as demonstrated in this paper; *iii)* TI-PSM scheme retains the complete avoidance of inter-symbol interference (ISI) and inter-channel interference (ICI) by transmitting the same modulated symbol from all groups; and *iv)* Only one RF chain is required in TI-PSM scheme, although multiple transmit antennas are activated simultaneously as all active antennas transmit the same modulated symbol. As such, these features reveal that the proposed TI-PSM scheme is capable of improving the reliability of the wireless channels since replicas of the transmitted information symbols are provided to the receiver.

The rest of the present paper is organized as follows. Section II introduces PSM scheme and presents the system model of the proposed TI-PSM scheme. The BER performance of TI-PSM system is presented in Section III. Finally, conclusions are summarized in Section IV.

*Notation:* The upper- and lower-case boldface letters denote matrices and vectors, respectively. $\|\mathbf{A}\|$ denotes the Frobenius norm operation of $\mathbf{A}$, $\lfloor \cdot \rfloor$ stands for the floor operation that flooring a real number to the nearest smallest integer, $\binom{a}{b}$ denotes the combinations without repetition of $a$ objects taken $b$ at a time, and $\mathbb{C}^{m \times n}$ denotes a complex matrix having the size of $m \times n$.

## II. SYSTEM MODEL

In this paper, time-indexed parallel spatial modulation scheme using antenna grouping is proposed as depicted in Fig. 1, where information bits are conveyed through indexing of multiple entities simultaneously. More specifically, indexing is carried out in spatial and time domains in PSM scheme.

In PSM scheme that considered in this paper, the available transmit antennas $N_t > 2$ are equally divided into $G > 1$ groups, where each group includes $N_{tg} \geq 2$ antennas. The information bit sequence to be conveyed at each time slot is divided into $G + 1$ parts. The first $G$ parts are to perform SM in parallel for transmitting the same data symbol from all active antennas, where only a single antenna is activated in each group, and the last part is to map the $M$-PSK/QAM complex constellation symbol. In other words, $G \log_2(N_{tg})$ bits are used to determine the active antennas in $G$ groups, and $\log_2(M)$ bits are modulated and transmitted from the active transmit antennas, simultaneously. Therefore, the spectral efficiency of PSM scheme in bits per channel use (bpcu) can be expressed as:

$$\eta_{PSM} = G \log_2(N_{tg}) + \log_2(M). \quad (1)$$

One of the main advantages of this scheme is that the number of groups can be any integer number, while the total number of transmit antennas can be any even number. This shows that the proposed PSM scheme can overcome the constraint of SM scheme where the number of transmit antennas has to be a power of two. For example, $G$ can be equal to 3 or 5 groups, and $N_t$ can be equal to 6 or 10 transmit antennas as illustrated in Figs. 2 and 3 in Section III.

In this paper, a TI-PSM scheme is proposed, where time slots and antennas are indexed simultaneously. The proposed TI-PSM scheme has $N_t$ transmit antennas along with $G \leq N_t/2$ active antennas, and information bits are conveyed through indexing time-slots and active antennas, besides $M$-QAM or PSK symbols. In this scheme, the time is divided into multiple data frames, and each frame consists of $T + L - 1$ time slots, where $T$ is the length of the corresponding data part of the frame in number of time slots, $L$ is the number of taps in the multipath frequency-selective channel, and $L - 1$ is the number of time slots that used for transmitting cyclic prefix (CP) [12]. Time indexing in TI-PSM is done by selecting only $T_a$ time slots out of $T$ time slots in a frame to convey $\left\lfloor \log_2 \binom{T}{T_a} \right\rfloor$ information bits, which are called 'time index bits'. The $T$-length pattern of active and inactive status of the time slots is called a time-slot activation pattern (TAP). In each active time slot, $G$ active antennas are activated to transmit the signal symbol based on $G \log_2(N_{tg})$ bits, with $\log_2(N_{tg})$ bits from each group. These bits are called 'antenna index bits'. Since there are $T_a$ out of $T$ signaling time slots in each data frame in the proposed TI-PSM transmission technique, a number of $GT_a \log_2(N_{tg})$ information bits are transmitted through antenna indexing in each frame instead of transmitting only $T_a \log_2(N_t)$ bits in TI-SM transmission scheme. In addition, $T_a \log_2(M)$ information bits are conveyed by the conventional modulation symbols in each frame. By contrast, the achieved rate of TI-SM scheme (in bpcu) is written as [12]

$$\eta_{TI-SM} = \frac{1}{T + L - 1}\left\{ \left\lfloor \log_2 \binom{T}{T_a} \right\rfloor + T_a \log_2(N_t M) \right\}, \quad (2)$$

while the achieved rate of the proposed TI-PSM scheme can be expressed as

$$\eta_{TI-PSM} = \frac{1}{T + L - 1}\left\{ \left\lfloor \log_2 \binom{T}{T_a} \right\rfloor + T_a[G \log_2(N_{tg}) + \log_2(M)] \right\}. \quad (3)$$

Similar to the case of TI-SM scheme introduced in [3], an $N_{tg} \times 1$ PSM signal vector gets transmitted from each group in each active time slot, and nothing is conveyed in inactive time slots [12]. The PSM signal set of each group can be expressed as

$$\mathbb{X}_{PSM} = \{\mathbf{x}_{n,m}: n = 1, 2, \ldots, N_{tg}, m = 1, 2, \ldots, M\}$$

$$\mathbf{x}_{n,m} = \begin{bmatrix} 0 \ldots 0 & \underbrace{x_m}_{n\text{th coordinate}} & 0 \ldots 0 \end{bmatrix}^T, x_m \in M. \quad (4)$$

For example, if $N_t = 12$, $G = 3$ and $M = 2$ for Binary PSK, then PSM signal set of each group is written as

$$\mathbb{X}_{PSM} = \left\{ \begin{bmatrix}1\\0\\0\\0\end{bmatrix}, \begin{bmatrix}-1\\0\\0\\0\end{bmatrix}, \begin{bmatrix}0\\1\\0\\0\end{bmatrix}, \begin{bmatrix}0\\-1\\0\\0\end{bmatrix}, \begin{bmatrix}0\\0\\1\\0\end{bmatrix}, \begin{bmatrix}0\\0\\-1\\0\end{bmatrix}, \begin{bmatrix}0\\0\\0\\1\end{bmatrix}, \begin{bmatrix}0\\0\\0\\-1\end{bmatrix} \right\}. \quad (5)$$

In this corresponding, the signal set of the proposed TI-PSM scheme can be expressed as

$$\mathbb{X}_{TI-PSM} = \{\mathbf{s} = [\mathbf{s}_1^T \mathbf{s}_2^T \ldots \mathbf{s}_T^T]^T : \mathbf{s}_i \in \mathbb{X}_{PSM} \cup \mathbf{0},$$
$$\|\mathbf{s}\|_0 = T_a \text{ and } \mathbf{t}^\mathbf{s} \in \mathbb{T}\}, \quad (6)$$

where $\mathbf{s}_1, \mathbf{s}_2, \ldots, \mathbf{s}_T$ denote the transmitted signal vectors in $T$ time slots, $\mathbf{0}$ is an $N_{tg} \times 1$ zero vector, $\mathbb{T}$ represents the set of all valid TAPs, and $\mathbf{t}^\mathbf{s}$ denotes the TAP that corresponding to the signal vector $\mathbf{s}$.

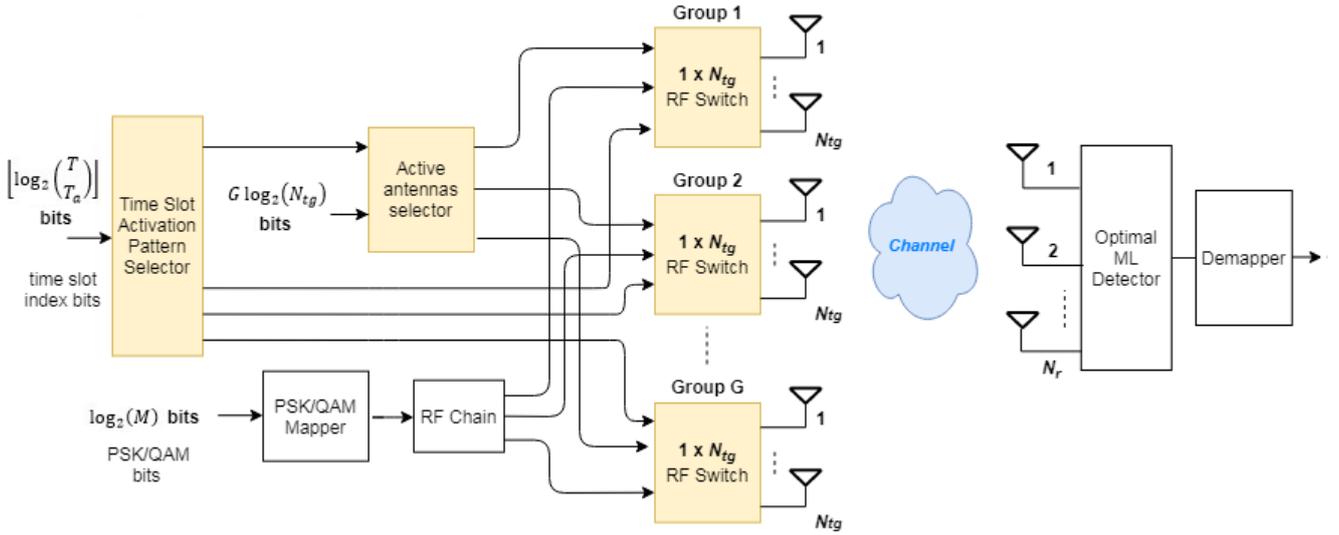

Fig. 1. System model of the proposed TI-PSM scheme with transmit antenna grouping using single RF chain.

It is worth mentioning that the size of TI-PSM signal set is $|\mathbb{X}_{TI-PSM}| = 2^{\lfloor \log_2 \binom{T}{T_a} \rfloor} (N_{tg}{}^G M)^{T_a}$, while in TI-SM scheme, $|\mathbb{X}_{TI-SM}| = 2^{\lfloor \log_2 \binom{T}{T_a} \rfloor} (N_t M)^{T_a}$. For example, if $T = 4$, $T_a = 2$, $M = 8$, and $N_t = 8$ with $G = 4$ two-antenna groups, then $|\mathbb{X}_{TI-PSM}| = 65536$. In TI-PSM scheme, an $TN_t \times 1$ signal vector from $\mathbb{X}_{TI-PSM}$ is conveyed over $T$ time slots in a frame.

After removing the CP at the receiver, the received signal vector is written as

$$\mathbf{y} = \mathbf{Hs} + \mathbf{n} \quad \in \mathbb{C}^{TN_r \times 1}, \quad (7)$$

where $N_r$ denotes the number of receive antennas, $\mathbf{n} \in \mathbb{C}^{TN_r \times 1}$ denotes a complex zero-mean additive white Gaussian noise (AWGN) vector with a variance of its elements being $\sigma_n^2$ per dimension at the receiver input, $\mathbf{n} \sim \mathcal{CN}(\mathbf{0}_{N_r}, \sigma_n^2 \mathbf{I}_{N_r})$, $\mathbf{I}_{N_r}$ is the $N_r \times N_r$ identity matrix, and $\mathbf{H} \in \mathbb{C}^{TN_r \times TN_t}$ is the equivalent channel matrix that written as

$$\mathbf{H} = \begin{bmatrix} \mathbf{H}_0 & \mathbf{0} & \mathbf{0} & \cdots & \mathbf{H}_{T_a-1} & \cdots & \mathbf{H}_1 \\ \mathbf{H}_1 & \mathbf{H}_0 & \mathbf{0} & \cdots & \mathbf{0} & \cdots & \mathbf{H}_2 \\ \vdots & & & \vdots & & & \vdots \\ \mathbf{H}_{T_a-1} & \mathbf{H}_{T_a-2} & \vdots & \mathbf{H}_0 & \mathbf{0} & \cdots & \mathbf{0} \\ \mathbf{0} & \mathbf{H}_{T_a-1} & \vdots & \mathbf{H}_1 & \mathbf{H}_0 & \cdots & \mathbf{0} \\ \vdots & & & \vdots & & \ddots & \\ \mathbf{0} & \mathbf{0} & \cdots & \cdots & \cdots & \cdots & \mathbf{H}_0 \end{bmatrix}, \quad (8)$$

where $\mathbf{H}_i \in \mathbb{C}^{N_r \times N_t}$ is the channel matrix of the $i$-th active time slot, $i = 0, 1, \ldots, T_a - 1$. It is assumed that the $k$-th column of $\mathbf{H}_i$ has zero mean and unit variance per dimension, $\mathbf{h}_i^k \sim \mathcal{CN}\left(\mathbf{0}_{N_r}, \frac{1}{L} \mathbf{I}_{N_r}\right)$. Without loss of generality, a flat Rayleigh fading channel with single path ($L = 1$) is assumed in this paper.

At the receiver of the proposed TI-PSM scheme, the maximum likelihood (ML) optimum detection is written as

$$[\widehat{\mathbf{A}}_k, \hat{\mathbf{s}}] = \arg\min_{\mathbf{S} \in M^{T_a}} \|\mathbf{y} - \mathbf{HS}\|^2, \quad (9)$$

where $\widehat{\mathbf{A}}_k$ is the index of the $k$-th antenna activation matrix that has $T_a$ columns to give the activation patterns for the $T_a$ active time slots, and $\mathbf{S}$ is the signal constellation space. The detected index $\widehat{\mathbf{A}}_k$ and the estimated symbol bits $\hat{\mathbf{s}}$ are then used to retrieve the original transmitted data bits.

At the receiver of practical systems, the imperfect channel estimation should be taken into consideration in order to decode the transmitted information. The estimate of $\mathbf{H}$ is denoted in this paper by $\widetilde{\mathbf{H}}$, and both are assumed to be jointly ergodic and stationary processes, and an orthogonality between the CEEs and the channel estimate is also assumed. For the imperfect channel state information (CSI) scenario, the estimated channel matrix $\widetilde{\mathbf{H}}$ with error is obtained as [13]

$$\widetilde{\mathbf{H}} = \mathbf{H} - \mathbf{E}, \quad (10)$$

where $\mathbf{E} \in \mathbb{C}^{TN_r \times TN_t}$ represents the CEE matrix with independent and identically distributed (i.i.d.) entries that having zero mean and variance $\sigma_e^2$, $\mathbf{E} \sim \mathcal{CN}(\mathbf{0}_{N_r}, \sigma_e^2 \mathbf{I}_{N_r})$. This variance captures the channel estimation quality, and it can be chosen appropriately depending on the channel estimation and dynamics schemes. For orthogonal pilot designs, it is assumed in this paper that the estimation error reduces linearly as the number of pilots increases [13, 14], thus the error variance $\sigma_e^2$ is kept equal to the noise variance $\sigma_n^2$.

Due to the sparsity of the codewords of the proposed TI-PSM scheme, as well as the high ML receiver complexity that can be written as $\mathcal{O}\left(\frac{T^{T_a+1}(N_{tg}{}^G M)^{T_a} N_r}{T_a^{T_a-1}}\right)$, which increases exponentially with the increase of $T_a$, further investigations with advanced sparse recovery algorithms and signal processing methods such as compressive sensing based detection techniques [15] may be employed in order to reduce the computational complexity of the proposed scheme.

## III. PERFORMANCE RESULTS AND COMPARISONS

In this section, the error performance of the proposed TI-PSM scheme that based on transmit antenna grouping is presented and compared to the performance of the conventional PSM and TI-SM schemes for the same achieved rate. Throughout the simulations, a flat Rayleigh fading channel is assumed with $L = 1$, and the number of receive antennas is kept constant, $N_r = 4$, in all schemes. For time-indexed schemes, it is assumed that $T_a = 2$ active time slots are used for transmission out of $T = 4$ signaling time slots.

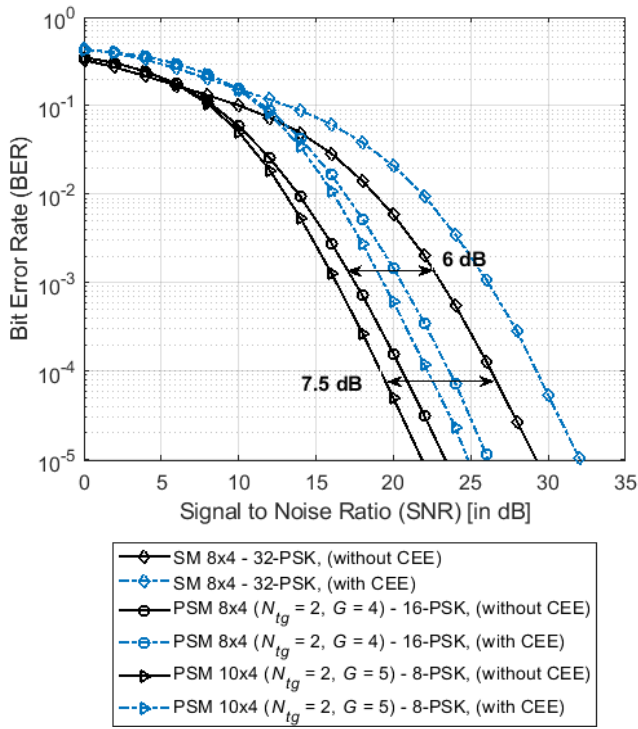

Fig. 2. BER performance of SM (diamond marker) and PSM schemes (circle and arrow markers) for 8 bpcu using ML detection without CEEs (solid lines) and with CEEs (dashed lines).

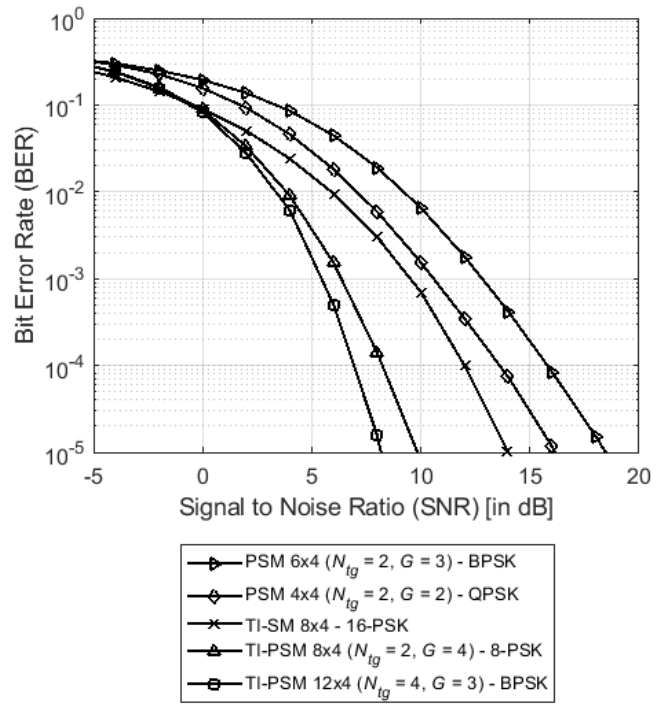

Fig. 3. BER performance of PSM, TI-SM and TI-PSM schemes for 4 bpcu using ML detection with assuming that a perfect CSI is available at the receiver. It is also assumed in time-indexed schemes that $T_a = 2$ active time slots are used for transmission out of $T = 4$ signaling time slots.

Fig. 2 compares the BER performance of the conventional SM and PSM schemes for eight bits transmission, and shows the effect of CEEs on the BER performance with assuming that an imperfect CSI is available at the receiver. This figure shows that PSM $8 \times 4$ with 4 groups and PSM $10 \times 4$ with 5 groups outperform the conventional SM $8 \times 4$ scheme by almost 6 dB and 7.5 dB, respectively. This is mainly because SM scheme requires a higher modulation order than that of PSM schemes to achieve the same spectral efficiency. In other words, the required modulation order in PSM systems is lower than that of SM systems for the same achieved rate, and it decreases as the number of groups and transmit antennas increases, hence yielding a significant BER performance improvement. It is also clear from Fig. 2 that a performance drop of about 3 dB is shown with CEEs in SM and PSM schemes as compared to the perfect CSI case without CEEs. Additionally, Fig. 2 shows that PSM schemes with imperfect CSI outperform SM scheme with imperfect and perfect CSI.

The BER performance of the proposed TI-PSM scheme is numerically investigated and compared to the performance of PSM and TI-SM schemes in Fig. 3 for four bits transmission with assuming that a perfect CSI is available at the receiver. It can be clearly seen from this figure that TI-PSM $8 \times 4$ with 4 groups and TI-PSM $12 \times 4$ with 3 groups outperform TI-SM $8 \times 4$ scheme by almost 4 dB and 5.5 dB at the BER of $10^{-5}$, respectively. It is also shown from Fig. 3 that TI-PSM $8 \times 4$ scheme with 4 groups outperforms PSM $6 \times 4$ with 3 groups and PSM $4 \times 4$ with 2 groups by approximately 6.5 dB and 8.5 dB at the BER of $10^{-5}$, respectively. In addition, TI-PSM $12 \times 4$ scheme with 3 groups outperforms PSM $6 \times 4$ with 3 groups and PSM $4 \times 4$ with 2 groups by approximately 8 dB and 10 dB at the BER of $10^{-5}$, respectively. This shows that significant improvements are achieved in time-indexed PSM schemes in which indexing is done in time slots and antennas besides the conventional modulation scheme. It is mainly because a lower modulation order is required in TI-PSM schemes as compared to TI-SM scheme for the same achieved rate, even with the same number of transmit antennas. This is because of the extra time-index bits that are exploited by transmission through indices of the used time slots in PSM schemes.

A comparison between the BER performance of TI-SM and TI-PSM schemes for 4 bpcu with perfect and imperfect CSI knowledge at the receiver is shown in Fig. 4. It is clear from this comparison that a performance degradation of about 5.5 dB is observed in all time-indexed schemes with the presence of CEEs over the same schemes with perfect channel knowledge at the receiver. On the other hand, another comparison between PSM and TI-PSM schemes for 4 bpcu is depicted in Fig. 5 where it showed that the performance drop in PSM schemes is about 3 dB as compared to the same schemes with assuming that a perfect CSI is available at the receiver. These results show that TI-SM and TI-PSM schemes where indexing is performed in time domain besides indexing in the spatial domain provide higher performance degradation than that of PSM schemes with the presence of CEEs. In addition, Fig. 5 shows that TI-PSM schemes with CEE outperform all PSM schemes with and without CEEs.

IV. CONCLUSION

In this paper, a new modulation scheme referred to as space-time parallel index modulation has been proposed in which transmit antennas and time slots are indexed simultaneously. This modulation scheme takes the advantage of PSM systems using antenna grouping, and tine-indexing approach to provide higher spectral efficiency and simple transceiver design by combining these techniques in a joint transmission mechanism, as well as the ICI avoidance in large-scale MIMO systems. The simulation results showed that indexing time slots in PSM schemes with single RF chain is a beneficial approach where significant BER performance improvements are achieved over TI-SM and PSM schemes for the same achieved rate. It has been also demonstrated that time-indexed schemes provide higher performance drop than that of the same schemes without time indexing when taking CEEs into account as compared to the same schemes with perfect CSI at the receiver. Furthermore, the significant improvements that are achieved make the proposed TI-PSM scheme a very efficient and highly suitable scheme for large-scale MIMO communications and 5G wireless networks.

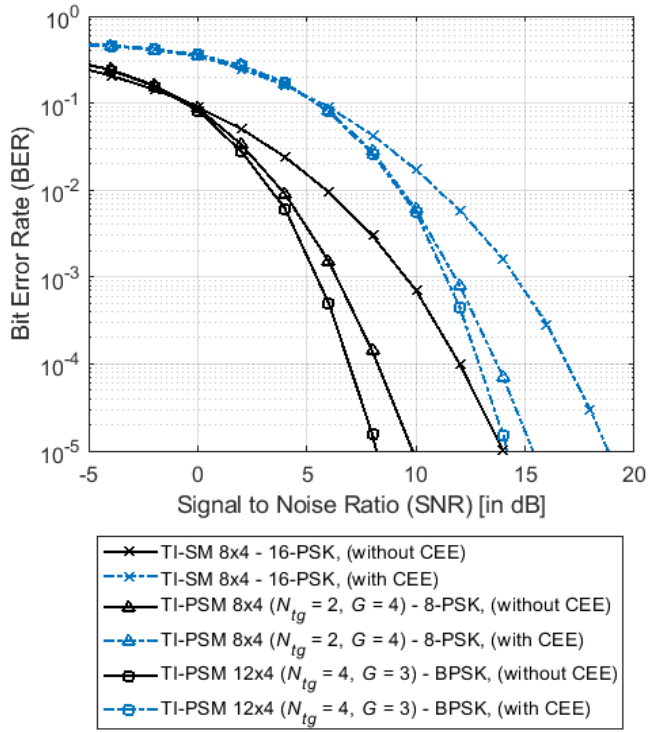
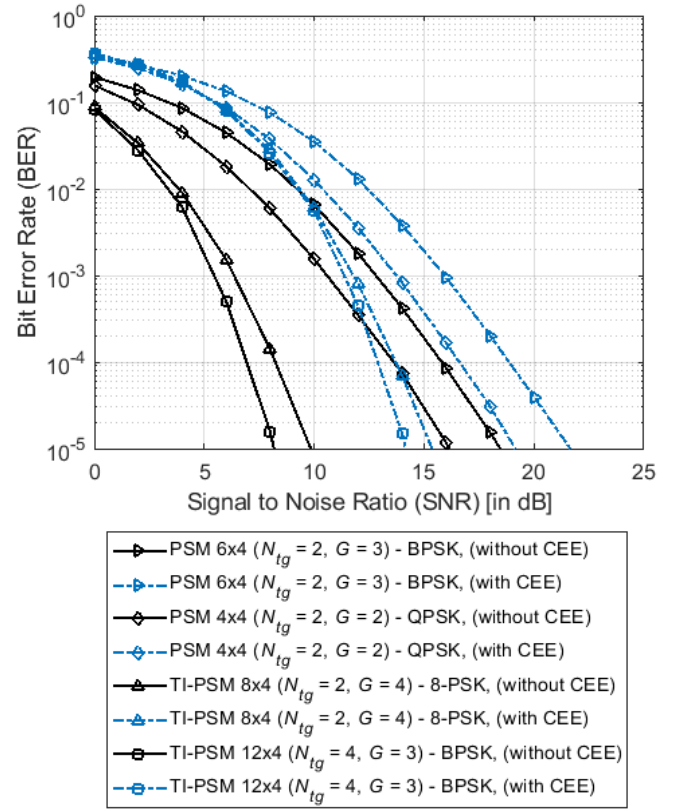

Fig. 4. BER performance of TI-SM and TI-PSM schemes for 4 bpcu using ML detection without CEEs (solid lines) and with CEEs (dashed lines). It is assumed that $T_a = 2$ active time slots are used for transmission out of $T = 4$ signaling time slots.

Fig. 5. BER performance of PSM and TI-PSM schemes for 4 bpcu using ML detection without CEEs (solid lines) and with CEEs (dashed lines). It is assumed in TI-PSM schemes that $T_a = 2$ active time slots are used for transmission out of $T = 4$ signaling time slots.